\documentclass[prl,twocolumn,superscriptaddress,nofootinbib, floatfix]{revtex4-1}

\usepackage{lipsum}
\usepackage{braket}
\usepackage{graphicx}
\usepackage{dcolumn}
\usepackage{bm}
\usepackage{amsmath}
\usepackage{mathtools}
\usepackage{subfigure}
\usepackage{color}
\usepackage{xcolor}
\usepackage{verbatim}
\usepackage{textcomp} 

\newcommand{\abs}[1]{\lvert #1 \rvert}

\begin{document}

\title{Near ground-state cooling of two-dimensional trapped-ion crystals with more than 100 ions}
\author{Elena Jordan}
\email[]{elena.jordan@nist.gov}
\affiliation{Time and Frequency Division, National Institute of Standards and Technology, Boulder, Colorado 80305, USA}

\author{Kevin A. Gilmore}
\affiliation{Time and Frequency Division, National Institute of Standards and Technology, Boulder, Colorado 80305, USA}
\affiliation{JILA, NIST, and Department of Physics, University of Colorado Boulder, Boulder, Colorado 80309, USA}

\author{Athreya Shankar}
\affiliation{JILA, NIST, and Department of Physics, University of Colorado Boulder, Boulder, Colorado 80309, USA}

\author{Arghavan Safavi-Naini}
\affiliation{JILA, NIST, and Department of Physics, University of Colorado Boulder, Boulder, Colorado 80309, USA}

\author{Justin G. Bohnet}
\affiliation{Time and Frequency Division, National Institute of Standards and Technology, Boulder, Colorado 80305, USA}

\author{Murray J. Holland}
\affiliation{JILA, NIST, and Department of Physics, University of Colorado Boulder, Boulder, Colorado 80309, USA}

\author{John J. Bollinger}
\affiliation{Time and Frequency Division, National Institute of Standards and Technology, Boulder, Colorado 80305, USA}

\date{\today}

\begin{abstract}
We study, both experimentally and theoretically, electromagnetically induced transparency cooling of the drumhead modes of planar 2-dimensional arrays with up to $N\approx 190$ Be${}^+$ ions stored in a Penning trap. Substantial sub-Doppler cooling is observed for all $N$ drumhead modes.  Quantitative measurements for the center-of-mass mode show near ground state cooling with motional quantum numbers of $\bar{n} = 0.3\pm0.2$ obtained within 200 \textmu{}s. The measured cooling rate is faster than that predicted by single particle theory, consistent with a quantum many-body calculation. For the lower frequency drumhead modes, quantitative temperature measurements are limited by apparent damping and frequency instabilities, but near ground state cooling of the full bandwidth is strongly suggested. This advancement will greatly improve the performance of large trapped ion crystals in quantum information and quantum metrology applications. 
\end{abstract}


\maketitle
Motivated by metrology and quantum informatic applications, along with the fundamental interest in controlling quantum degrees of freedom, the preparation of mechanical oscillators close to their quantum mechanical ground state has been an active pursuit for three decades. Early examples include cooling the high-frequency cyclotron motion of a single trapped electron \citep{Piel2003} by refrigeration (i.e.\  coupling it to a cold environment), and laser sideband cooling the lower frequency motion of single trapped ions \citep{Diedrich1989} and trapped neutral atoms \citep{Perrin1998}. More recently, single, high-Q modes of macroscopic mechanical oscillators have been cooled close to the ground state, either by refrigeration when the mode frequency was sufficiently high \citep{Hofheinz2010}, or through sideband cooling \citep{Teufel2011,Chan2011}.

Simultaneously ground-state cooling many modes of a macroscopic resonator or a large trapped-ion crystal remains a challenge. Traditional sideband cooling has been extended to cooling multiple modes of small ion crystals \citep{King1998, Stutter2018}, but does not scale well as the number of ions and modes increases. Electromagnetically induced transparency (EIT) cooling \citep{Morigi2000, Morigi2003, Roos2000, Lin2013, Xia2009} shows promise for greatly extending the number of modes or motional degrees of freedom that can be ground-state cooled, as was shown for the radial modes of a linear string of 18 ions \citep{Lechner2016}. 
Here we demonstrate experimentally and theoretically near ground state cooling for all the axial drumhead modes of two-dimensional crystals with more than 100 ions, giving us exquisite quantum control of a mesoscopic system. Aside from the intrinsic interest in preparing larger mesoscopic systems close to their quantum mechanical ground state, ground state cooling improves the application of large ion crystals for quantum computation and simulation \citep{Britton2010, Gaerttner2017} and for weak force sensing \citep{Gilmore2017}.

\begin{figure}[ht!]
    \centering
    \includegraphics[width=\columnwidth]{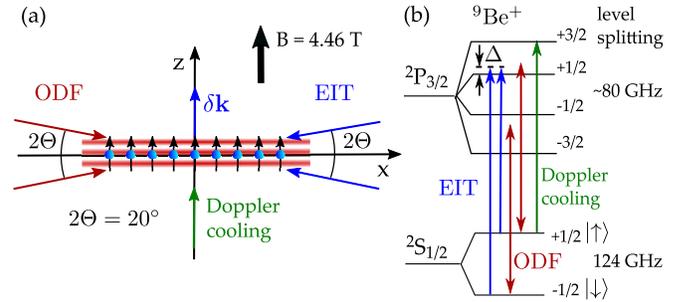}
    \caption{(a) Schematic laser setup for EIT cooling. The blue spheres represent the ions with their spins (arrows). The beams generating the spin-dependent optical-dipole force (ODF) (red) cross the ion plane at $\Theta = \pm 10^{\circ}$, the EIT cooling beams (blue) are counter-propagating relative to the ODF beams. The ODF beams interfere at the position of the ions and form a traveling wave potential (red fringes). (b) Level diagram of $\prescript{9}{}{}$Be$^{+}$ in the 4.46 T magnetic field. The two ODF beams couple to both hyperfine levels of the ground state and have a frequency difference that can be stepped across the drumhead mode frequencies.} 
    \label{EITsetup}
\end{figure}
In general, EIT cooling takes advantage of the phenomenon of coherent population trapping \citep{Lounis1992}. Two energetically lower lying states are coherently coupled by two lasers to the same excited state with equal detuning $\Delta$ (Fig.~\ref{EITsetup}). The couplings can be described by two excitation amplitudes. The interference between these excitation amplitudes leads to a Fano-shaped profile in the population of the excited state as a function of the frequency of a weak probe \citep{Lounis1992, Morigi2000, Morigi2003}. The Fano-shaped profile features both a zero in the absorption (dark state), where the excitation amplitudes interfere destructively, as well as a narrow absorption resonance which is on the higher frequency side of the zero for blue detuning ($\Delta>0$). For large blue detuning, the distance $\delta$ between the dark state and narrow resonance is given by the light shift due to the two lasers. By choosing the power of the laser beams so that $\delta$ is equal to the motional frequency, the motion subtracting sideband falls on the narrow absorption resonance and is strongly enhanced, while the motion adding sideband is suppressed. Thus, the ions predominantly scatter a photon while simultaneously losing a quantum of motion, thereby cooling the system \citep{Morigi2000}. 

Our experimental setup is shown in Fig.\ \ref{EITsetup}. We use Doppler laser cooling in a Penning trap, which employs a strong, uniform magnetic field $(B=4.46\;\text{T})$ and static electric fields, to form a single-plane Coulomb crystal with $N\leq 200$ $\prescript{9}{}{}$Be$^{+}$ ions 
\cite{hydrides, Sawyer2014, Bohnet2016}. The spin-1/2 degree of freedom is the $\prescript{2}{}{S}_{1/2}$ ground-state valence electron spin $\ket{\uparrow} (\ket{\downarrow}) \equiv \ket{m_{s}=+1/2} (\ket{m_{s}=-1/2}) $. At the magnetic field of 4.46 T, the Zeeman splitting of this ground state is 124 GHz. Global spin rotations are driven by a resonant microwave source.

The ions are confined in the direction parallel to the magnetic field by a harmonic electrostatic potential characterized by a frequency (the center-of-mass (c.m.)\  frequency) $\omega _{\text{c.m.}}/(2\pi)=1.59 \;\text{MHz}$. The Doppler cooling limit for  $\prescript{9}{}{}$Be$^{+}$ of approximately 0.4 mK corresponds to a mean phonon occupation number for the c.m.\ mode of $\bar{n} = 4.6$ \citep{Torrisi2016}. In a direction perpendicular to the magnetic field the ions are confined by the Lorentz force generated by the rotation through the magnetic field. The rotation frequency is controlled with a `rotating wall' potential and set to produce a single plane crystal, typically  $\omega _{\text{rot}}/(2\pi)= 180.0$ kHz. For these single-plane crystals, motion along the trap axis (i.e.\ parallel to \textbf{z}), is described by $N$ axial modes, which we refer to as the drumhead modes \citep{Sawyer2014,shankar2018}. Furthermore, there are $2N$ `in-plane modes' associated with motion in the plane of the crystal (i.e.\ orthogonal to \textbf{z}) \cite{Wang2013}. Only the drumhead modes are utilized for applications such as quantum simulation and sensing, and we focus on EIT cooling of the drumhead modes in this Letter.

 For EIT cooling, the $\ket{\uparrow}$ and $\ket{\downarrow}$ states are coherently coupled to the $^2P_{3/2}$ $\ket{m_J = +1/2}$ excited state using two lasers with 313 nm wavelength. The two lasers generating the EIT interaction are phase-locked with a frequency offset equal to the spin-flip frequency (124 GHz) \cite{suppMat}, and intersect with the ion crystal at $\pm10^{\circ}$ angles. The collimated beams have a $1/e^2$ diameter of 1 mm providing an approximately uniform intensity over the entire crystal  (diameter $< 250$ \textmu{}m for $N <200$). To minimize the interaction of the EIT beams with the in-plane motion, the EIT $\Delta \mathbf{k}$ vector is aligned normal to the crystal plane (parallel to the magnetic field z axis), with an estimated misalignment of $<0.2^{\circ}$. Because the EIT beams are not normal to the single plane crystal, the ion crystal rotation  produces a time-varying Doppler shift, which can be several hundred MHz for ions on the crystal boundary, and leads to an effective modulation of the detuning $\Delta$. We chose $\Delta$ such that even with the largest Doppler shift the lasers are still effectively blue detuned \citep{shankar2018}, typically $\Delta = 400$~MHz. We adjust the powers of the EIT beams so that the light shift is equal to the c.m.\  mode frequency, and the two Rabi frequencies are equal in order to maximize the cooling rate \citep{Morigi2003}.

Complications such as the time-varying Doppler shifts, insufficient separation of electronic and motional timescales \citep{shankar2018}, and the simultaneous cooling of a large number of ions interacting through many modes demand careful numerical modeling of the potential efficacy of EIT cooling in a Penning trap. Encouragingly, theory \citep{shankar2018} predicts the possibility of near ground-state cooling for all the drumhead modes despite these challenges. 
\begin{figure}[t]
    \centering
    \includegraphics[width=\columnwidth]{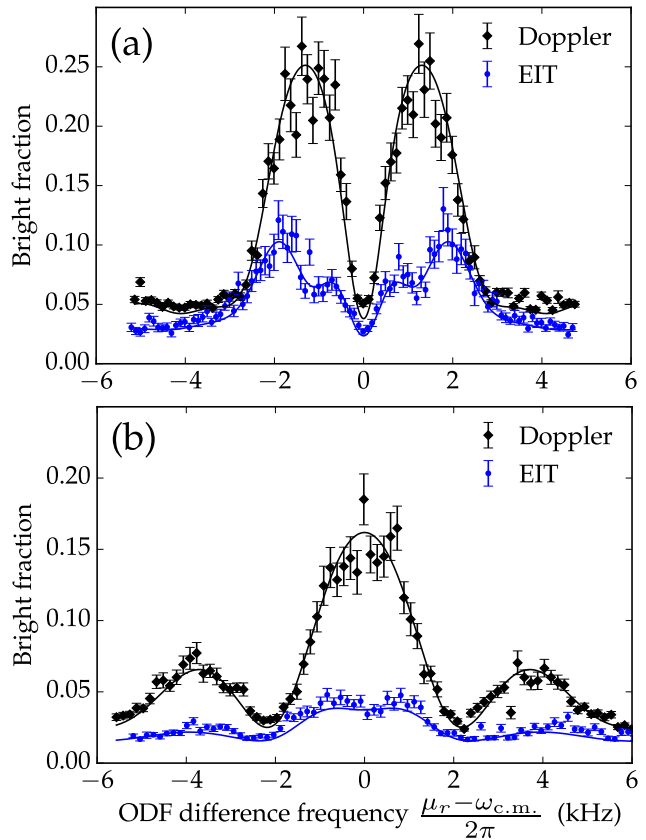}
    \caption{ \label{COMmode}a) Temperature measurement of the c.m.\ mode at frequency $\omega_{\text{c.m.}} = 2\pi \times 1.59$ MHz for a crystal with 158$\pm$10 ions. The black diamonds are the measured fraction of ions in the $\ket{\uparrow}$ state after Doppler cooling only, and the blue dots after Doppler cooling followed by 200 \textmu{}s of EIT cooling. The solid lines are least squares fits of Eq.\ (\ref{eqPup}) to the data. The ODF interaction time was $2\tau= 500$ \textmu{}s. The fitted mean c.m.\ mode occupations are $\bar{n}_{\text{Dop}} = 5.6\pm1.1$ after Doppler cooling only and $\bar{n}_{\text{EIT}} = 0.28\pm0.18$ after EIT cooling. The increased background for the Doppler cooling is due to a larger measured $\Gamma$ resulting from the weak Lamb-Dicke confinement. b) Same as a) but with a $\pi$ phase shift of the second ODF pulse \cite{suppMat}. Here a shorter ODF interaction time of $2\tau= 300$ \textmu{}s was used. The fits result in $\bar{n}_{\text{Dop}} = 5.1\pm0.8$ and $\bar{n}_{\text{EIT}} = 0.29\pm0.17$.}
\end{figure}

To measure the temperature of the drumhead modes, we couple the axial motion of the ions with their internal spin degree of freedom using a spin-dependent optical dipole force (ODF). The ODF is generated by two interfering off-resonant laser beams with beatnote frequency $\mu _r$ leading to a  traveling wave potential gradient along the z-direction. The resulting coupling is described by a Hamiltonian of the form $\hat{H} = F \cos(\mu _r t)  \sum_{j=1}^N \hat{z}_j \hat{\sigma}_j^z$, where $F$ is the ODF amplitude, and $\hat{z}_j$ and $\hat{\sigma}_j^z$ are the position operator and the Pauli spin matrix for ion $j$, respectively. 
When the frequency $\mu _r$ matches a drumhead mode frequency $\omega _i$, spin dephasing proportional to the amplitude of the motion occurs \citep{Gilmore2017}. To measure this spin dephasing, we use a Ramsey-style sequence \cite{suppMat}. The ions are prepared in the $\ket{\uparrow}$ state, a resonant microwave $\pi/2$ pulse rotates the spins around the y axis to align with the x axis, and the ODF produces spin precession for an interaction time of $2\tau$. Midway through the ODF interaction, the spin precession is interrupted by a $\pi$ pulse, implementing a spin-echo \cite{suppMat}. A final $\pi/2$ pulse is applied that brings the ions to the $\ket{\downarrow}$ state, if no dephasing occurred. Spin dephasing leads to a finite $\ket{\uparrow}$ state probability (denoted the bright fraction), which we measure through state-dependent resonance fluorescence on the Doppler cooling transition. The method is described in detail in Ref.\ \citep{Sawyer2012}. 

Figure \ref{COMmode} shows measurements of the bright fraction with the spin-echo sequence as the ODF difference frequency $\mu_r$ is stepped across the c.m.\ mode frequency $\omega_{\text{c.m.}}$.  A clear decrease in the bright fraction is observed when Doppler cooling is followed by EIT cooling, indicating a decrease in dephasing due to the lower c.m.\ mode temperature. To extract the mean c.m.\ mode occupation $\bar{n}$, we fit to an analytical expression \cite{suppMat} for the $\ket{\uparrow}$ state probability
\begin{equation}
P(\ket{\uparrow}) = \frac{1}{2}\left[1 - \exp\left(-2\Gamma \tau\right)C_{\text{ss}}C_{\text{sm}} \right],
\label{eqPup}
\end{equation}
where the coefficients $C_{\text{ss}} = (\cos(4J))^{N-1}$ and  $C_{\text{sm}}=\exp\left(-2\abs{\alpha}^2(2\bar{n}+1)\right)$  describe the phonon-mediated spin-spin interaction, and the dephasing that arises from spin-motion coupling, respectively. Here, $N$ is the number of $\prescript{9}{}{}$Be$^{+}$ ions and $2\tau$ is the total ODF interaction time. The spin-dependent displacement amplitude $\alpha$ and spin-spin coupling $J$ are functions of $\tau$, the spin-echo $\pi$-pulse duration $t_{\pi}$, the optical dipole force amplitude $F$ and the frequencies $\mu _r$ and $\omega_{\text{c.m.}}$ \cite{suppMat}. We determine $F$ from measurements of the mean-field spin precession \cite{Britton2010}. The decoherence rate $\Gamma$ is mainly due to spontaneous emission and is measured with the same spin-echo sequence but with the ODF beat note $\mu_r$ tuned far from any drumhead mode frequencies so that $C_{\text{ss}}=C_{\text{sm}}=1$.

Figure \ref{COMmode} also shows least-squares fits of Eq.\ (\ref{eqPup}) to the experimental measurements where $\omega _{\text{c.m.}}$ and $\bar{n}$ are free parameters. From Eq.\ (\ref{eqPup}), the observed signal will include both a temperature-dependent spin-motion component $(C_\text{sm})$ and a spin-spin component $(C_\text{ss})$ that does not depend on the temperature. For the measurements after only Doppler cooling the signal is dominated by motion-induced dephasing, in contrast to the EIT cooling measurements where the spin-spin component dominates, giving rise to a distinct line shape. If the ODF phase is shifted by $\pi$ for the second arm of the sequence, the lineshape is altered and $\alpha$ and $J$ are adjusted according to the experimental sequence (Fig.\ \ref{COMmode}(b) and \cite{suppMat}).

All observed line shapes agree well with the theoretical predictions, enabling temperatures to be evaluated through fits to the model (Eq.\ (\ref{eqPup})). After EIT cooling we obtain consistent measurements of $\bar{n} = 0.28\pm0.18$ (without a phase shift) and  $\bar{n} = 0.29\pm0.17$ (with a $\pi$ phase shift), demonstrating near ground state cooling for the c.m.\ mode with greater than 100 ions.  For Doppler cooling only, occupancies of $\bar{n}=5.6\pm1.1$ (without a phase shift) and $\bar{n}=5.1\pm0.8$ (with a $\pi$ phase shift) are obtained, consistent with the Doppler cooling limit.

\begin{figure}[tb]
    \centering
    \includegraphics[width=0.9\columnwidth]{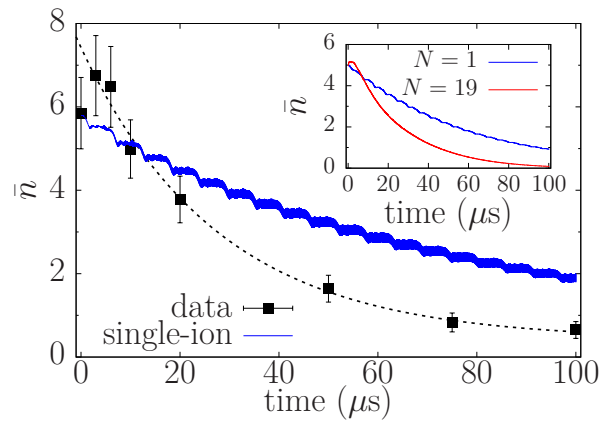}
  \caption{EIT cooling transient for the c.m.\ mode of a crystal with $N= 190\pm10$ ions. 
  The mean occupation of the c.m.\ mode $\bar{n}$ is plotted versus the EIT cooling time. The black dotted line is an exponential fit to the data. The blue curve shows the average cooling transient computed for single ions. 
  To approximately incorporate the radial crystal structure, we compute single-ion transients for each distance in the crystal from the trap center, and then average by weighting each transient by the number of ions at that radius. The thickness of the curve accounts for a 10\% uncertainty in the powers of the EIT lasers.  Inset: Simulated c.m.\ mode cooling transients for a single ion (blue) and a $19$-ion crystal (red). The cooling rate increases with ion number. }
  \label{coolingrate}
\end{figure}

 To determine a cooling rate for the c.m.\  mode, we measured the c.m.\ mode occupation $\bar{n}$ for increasing durations of EIT cooling. Figure~\ref{coolingrate} shows measurements obtained with a $2\tau=300$ \textmu{}s ODF interaction time and the sequence employed in Fig.\ \ref{COMmode}(b). The measured cooling transient can be well described by an exponential with $1/e$ time of $\tau _{\text{cool}} = 27.6 \pm1.7$ \textmu{}s. The measured heating of the c.m.\ mode is negligible on this time scale \cite{suppMat}. 
 
 This measured cooling rate is faster than the average rate expected from $N$ independently cooled ions (blue curve in Fig.~\ref{coolingrate}). This observation is consistent with detailed numerical simulations of EIT cooling with smaller crystals, where the cooling rate of the c.m.\ mode is found to increase with $N$ \cite{shankar2018}. The inset of Fig.~\ref{coolingrate} shows simulated transients for the c.m.\ mode for a single ion and a $19$-ion crystal, demonstrating an increase in cooling rate with the number of ions. The simulations also reproduce the experimentally observed initial heating during the first few microseconds, as the $N=19$ curve in the inset shows. The initial heating is caused by transient internal transitions until the ions reach the approximate dark state.
 
\begin{figure}[t]
    \centering
    \includegraphics[width=0.95\columnwidth]{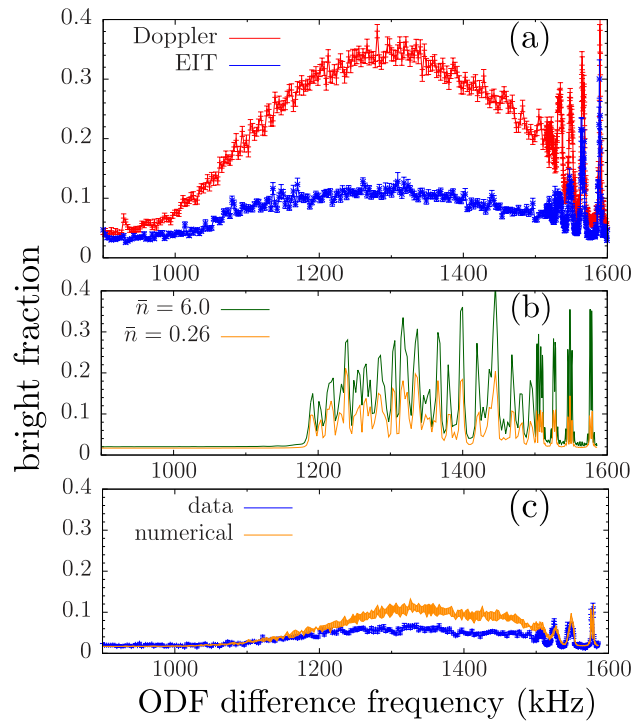}
  \caption{Scan over all the axial drumhead modes. a) The red points are data taken with Doppler cooling, blue points are data after Doppler and 200 \textmu{}s of EIT cooling for a crystal with $158\pm10$ ions. The significant reduction in the amplitude indicates a lower temperature of all the modes. b) Theoretical calculation of the bright fraction for a crystal with $N=79$ ions after Doppler cooling (green) and after Doppler and subsequent EIT cooling (orange) c) Comparison of the measured (blue) and the theoretically calculated (orange) bright fraction for a crystal with  $N=79\pm5$ ions. Here, the calculation took into account mode frequency instabilities leading to the loss of visibility of individual modes, and a 5\% uncertainty in the ODF amplitude $F$.}
  \label{allmodes}
\end{figure}

 The broad bandwidth of EIT cooling enables simultaneous cooling of all the drumhead modes without changing the experimental parameters. Figure \ref{allmodes}(a) shows spin-dephasing measurements as the ODF difference frequency is swept over the full bandwidth of axial modes of a crystal with $N= 158\pm 10$ ions after only Doppler cooling (red), and after Doppler cooling followed by 200 \textmu{}s of EIT cooling (blue). The significant reduction in the bright fraction over the entire bandwidth after EIT cooling is suggestive of substantial sub-Doppler cooling of all the drumhead modes.  
 
 A quantitative determination of the drumhead mode temperatures in Fig.\ \ref{allmodes}(a) is hindered by the large number of modes and, except for the highest frequency modes, the lack of resolved modes. Figures \ref{allmodes}(b) and (c) summarize an investigation of the drumhead mode temperatures for a smaller crystal with $N=79$ ions.  Figure \ref{allmodes}(b) shows a theoretical calculation of the bright fraction using Eq.\ (\ref{eqPup}) with an ODF interaction time $2\tau = 600$~\textmu{}s, both for Doppler cooling (green) and EIT cooling (orange) where all modes are assumed to have a thermal occupation with $\bar{n}=6$ and $\bar{n}=0.26$, respectively. The calculation assumes stable mode frequencies and indicates that for the parameters of the measurement the modes should be partially resolved.  This is to be contrasted with figure \ref{allmodes}(c), which shows an experimental scan (blue curve) over the drumhead mode bandwidth after EIT cooling of an $N=79\pm5$ ion crystal, for which we quantitatively measure the mean occupation of the c.m.\ mode to be $\bar{n}=0.26\pm0.38$.  As in Fig.\ \ref{allmodes}(a), a partially resolved mode structure is only observed for the highest frequency drumhead modes. 
 
 We model this lack of structure as being due to mode frequency fluctuations, which could arise from microscopic rearrangements of the ion crystal that are averaged over the many repetitions of an experiment, or possibly from damping due to coupling between modes. We expect frequency fluctuations from these sources to increase with decreasing wavelength, which in general corresponds to decreasing drumhead mode frequency. 
In Fig.\ \ref{allmodes}(c), we plot the theoretical bright fraction (orange) for the same crystal as in Fig.\ \ref{allmodes}(b) assuming all the modes are thermally occupied with $\bar{n}=0.26$, but now accounting for mode frequency fluctuations. By assuming Gaussian fluctuations that increase linearly from 1 kHz for the second-highest frequency mode to 80 kHz for the lowest frequency mode, the numerically computed bright fraction is  qualitatively similar to the experimentally observed spectrum (blue, Fig.\ \ref{allmodes}(c)).

In order to quantitatively determine the average motional quantum numbers of individual modes beyond the first few well-resolved modes, a detailed understanding of the sources, magnitudes, as well as timescales of the mode fluctuations is required.  However, the qualitative analysis of Fig.\ \ref{allmodes}(c) strongly suggests near ground state cooling for all the drumhead modes.  In the future improved modeling may enable a more quantitative analysis by including, for example, anharmonic corrections  
that can lead to coupling of the drumhead modes to the in-plane modes. 

In conclusion, we have demonstrated near ground-state EIT cooling of the entire bandwidth of drumhead modes of large planar ion crystals in a Penning trap. Quantitative measurements for the c.m.\ mode show mean occupations as low as $\bar{n}= 0.3\pm 0.2$ within 200 \textmu{}s, in good agreement with expectations from a numerical model \cite{shankar2018}. Further, the cooling is faster than predicted by single-ion calculations, documenting the many-body nature of the cooling process. Fast cooling rates and very low steady-state occupations enable EIT cooling to quickly initialize the axial modes to very low temperatures, thereby greatly improving the quality of quantum simulation and quantum metrology protocols. This result greatly increases the ion crystal size and number of phonon modes that can be cooled near to the ground state. Future work will study whether EIT cooling could be employed for ground state cooling for three-dimensional crystals with much larger numbers of ions.

\begin{acknowledgments}
We thank M.\ J.\ Affolter, Y.\ Lin, J.\ Cooper, R.\ J.\ Lewis-Swan, S.\ S.\ Kotler, and J.\ D.\ Teufel for stimulating discussions. We thank J.\ W.\ Britton and B.\ C.\ Sawyer for the help with preparatory work in the lab. We acknowledge the use of the Quantum Toolbox in Python (QuTiP) \cite{qutip, qutip2} for the exact single-ion numerical calculations presented in this paper.  This work was supported by NSF grants  PHY 1734006 and PHY 1820885, DARPA Extreme Sensing, the Air Force Office of Scientific Research grants FA9550-18-1-0319 and its Multidisciplinary University Research Initiative grant (MURI), Army Research Office grant W911NF-16-1-0576, JILA-NSF grant PFC-173400, and NIST. EJ gratefully acknowledges the Leopoldina Research Fellowship, German National Academy of Sciences grant LPDS 2016-15, and the NIST-PREP program. This manuscript is a contribution of NIST and not subject to U.S. copyright.
\end{acknowledgments}

\bibliographystyle{apsrev4-1}
\bibliography{EITcooling}

\begin{thebibliography}{28}
\bibitem [{hyd()}]{hydrides}%
  \BibitemOpen
  \href@noop {} {}\bibinfo {note} {In the room-temperature vacuum system,
  inelastic collisions with the H$_2$ background gas slowly converts Be$^{+}$
  ions to BeH$^{+}$ molecular ions that centrifugally separate to the radial
  edge of the crystal. For the measurements discussed here the number of
  impurity ions was less than $\leq 20$\,\% of the total $N$.}\BibitemShut
  {Stop}%
\end{thebibliography}

\end{document}


\title{Supplementary Material: Near ground-state cooling of two-dimensional trapped-ion crystals with more than 100 ions}
\author{Elena Jordan}
\email[]{elena.jordan@nist.gov}
\affiliation{Time and Frequency Division, National Institute of Standards and Technology, Boulder, Colorado 80305, USA}

\author{Kevin A. Gilmore}
\affiliation{Time and Frequency Division, National Institute of Standards and Technology, Boulder, Colorado 80305, USA}
\affiliation{JILA, NIST, and Department of Physics, University of Colorado Boulder, Boulder, Colorado 80309, USA}

\author{Athreya Shankar}
\affiliation{JILA, NIST, and Department of Physics, University of Colorado Boulder, Boulder, Colorado 80309, USA}

\author{Arghavan Safavi-Naini}
\affiliation{JILA, NIST, and Department of Physics, University of Colorado Boulder, Boulder, Colorado 80309, USA}

\author{Justin G. Bohnet}
\affiliation{Time and Frequency Division, National Institute of Standards and Technology, Boulder, Colorado 80305, USA}

\author{Murray Holland}
\affiliation{JILA, NIST, and Department of Physics, University of Colorado Boulder, Boulder, Colorado 80309, USA}

\author{John J. Bollinger}
\affiliation{Time and Frequency Division, National Institute of Standards and Technology, Boulder, Colorado 80305, USA}

\date{\today}

\pacs{}

\section*{Supplementary Material:\\ Near ground-state cooling of two-dimensional trapped-ion crystals with more than 100 ions}

\section{Experimental details}
For electromagnetically induced transparency (EIT) cooling of ${}^{9}$Be${}{}^{+}$ we used two commercial 1252 nm diode lasers, which we phase-locked in the infrared with an offset of 31 GHz. The 1252 nm light was amplified, and then frequency doubled, first to the visible (626 nm), and then to the UV (313 nm) via cavity-enhanced second-harmonic generation. 

The large Zeeman splittings produced by the 4.6 T magnetic field enable a nearly ideal 3-level EIT coupling between the ground state $\ket{2S_{1/2}, -1/2}$, $\ket{2S_{1/2}, +1/2}$ levels and a single excited state $\ket{2P_{1/2}, +1/2}$. The Rabi frequencies $\Omega_1$ for the $\ket{2S_{1/2}, -1/2}\rightarrow \ket{2P_{3/2}, +1/2}$ EIT laser beam and $\Omega_2$ for the $\ket{2S_{1/2}, +1/2}\rightarrow \ket{2P_{3/2}, +1/2}$ laser beam were separately adjusted by measuring the AC Stark shift on the 124 GHz transition with the laser frequency at $3.00\pm0.05$ GHz from resonance. Approximately equal Rabi frequencies $\Omega_1\approx\Omega_2$ were chosen to maximize the cooling rate. For a given blue detuning $\Delta$, $\Omega = \sqrt{\Omega_1^2 +\Omega_2^2}$ was chosen to give optimized cooling on the center of mass mode $\omega_z$ by satisfying $\delta=\omega_z$.  Here $\delta = \frac 12 (\sqrt{\Omega^2+\Delta^2}-|\Delta|) $ is the distance between the dark state and the narrow absorption resonance \cite{Morigi2003}.
This typically required tens of milliwatts of laser power per beam for a detuning of $\Delta = 400$ MHz and a $1/e^2$ diameter of 1 mm. The EIT beams were aligned so that $\delta\mathbf{k}$ was perpendicular to the plane of the ions within $\pm 0.2^{\circ}$. The alignment was done by aligning the EIT beams to be counter propagating to the ODF beams. The ODF beam $\delta \mathbf{k}$ vector was aligned perpendicular to the ion crystal plane following the procedure outlined in the Supplemental Material of \citep{Britton2010}. The EIT beam alignment was checked by confirming the lack of rotational sidebands on stimulated Raman transitions driven by the EIT beams when detuned from resonance by 3 GHz. 

\begin{figure}[h]
    \centering
    \includegraphics[width=0.63\columnwidth]{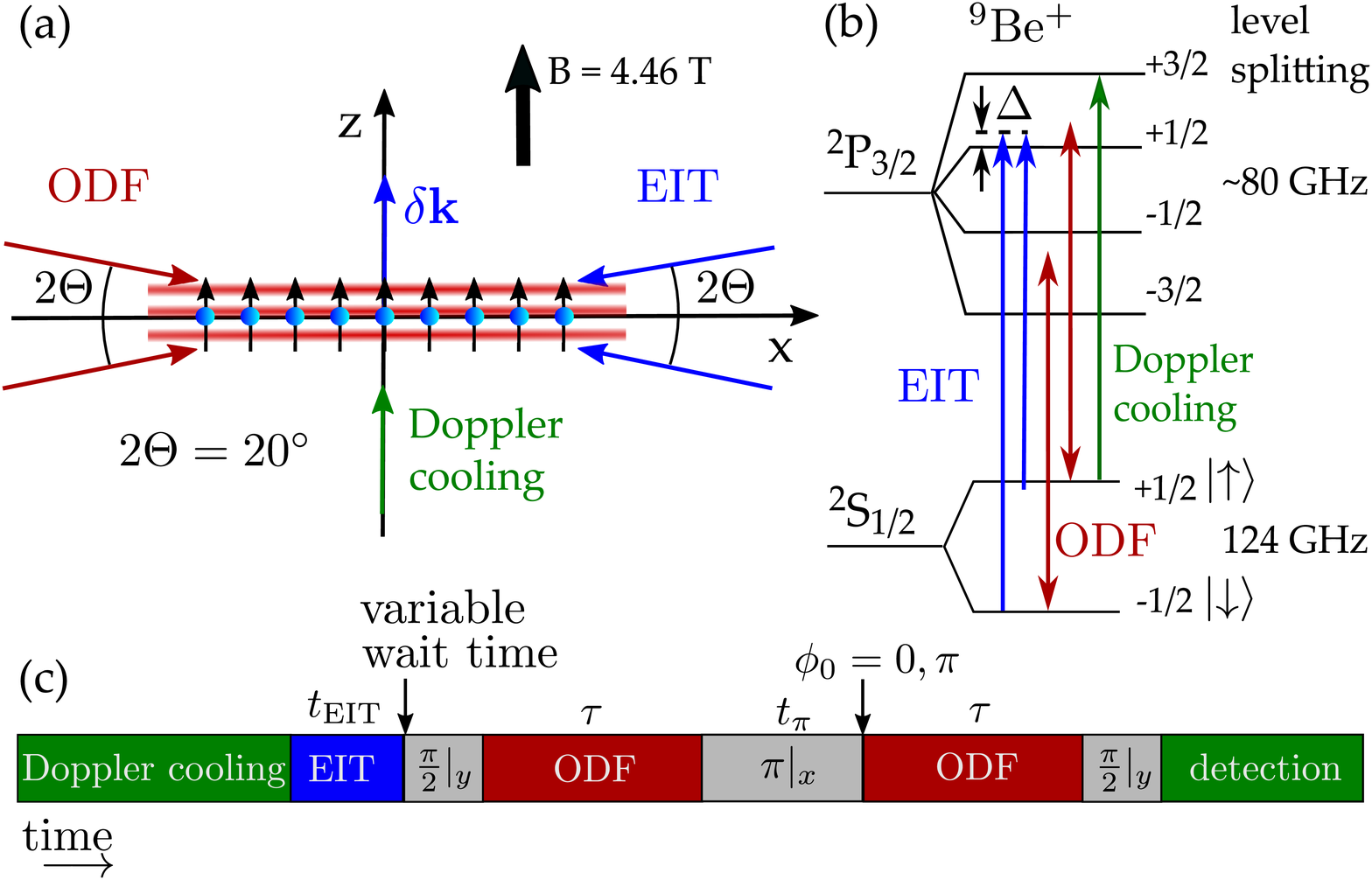}
  \caption{a) Experimental setup for EIT cooling. b) Level diagram of $\prescript{9}{}{}$Be$^{+}$ in the 4.6 T magnetic field of the Penning trap, with the lasers used in the experiment indicated with arrows. We show only the nuclear spin $m_I=3/2$ levels, which are prepared experimentally through optical pumping.  c) Experimental sequence. Pulse lengths not to scale. Doppler cooling and detection are much longer than indicated. The $\frac{\pi}{2}$ and $\pi$ pulses are implemented with resonant microwaves. The sequence was applied without a phase shift $\phi_0 = 0$ and with a $\pi$ phase shift $\phi_0 =\pi$ between the two ODF pulses. A variable wait time was inserted in the sequence to measure the heating rate.}
  \label{sequence}
\end{figure}

Figure \ref{sequence} shows the experimental sequence for EIT cooling and the temperature measurement. The ions are Doppler cooled for 10 ms. Then EIT cooling is applied for typically 200 \textmu{}s after which the ions are optically pumped into the $\ket{\uparrow}$ state by turning off the $\ket{2S_{1/2}, +1/2}\rightarrow \ket{2P_{3/2}, +1/2}$ EIT laser 300~\textmu{}s before the $\ket{2S_{1/2}, -1/2}\rightarrow \ket{2P_{3/2}, +1/2}$ EIT laser. To measure the motion-induced dephasing, a spin echo sequence is then applied. The duration for microwave-driven spin rotations by $\pi$ was $t_{\pi} \approx 80$ \textmu{}s. A spin-dependent optical dipole force was applied in the two arms of the spin echo sequence using laser beams detuned from resonance by $\approx 20$~GHz. For the detection of the spin state, the Doppler cooling laser was used. A long detection time of 2.5 ms was chosen to achieve a precision that is limited by spin projection noise \citep{Itano1993}. At each ODF difference frequency, the sequence was repeated 50 - 100 times. 

The heating rate of the c.m.\ mode was measured by inserting a variable wait time after the EIT cooling application (Fig.\ \ref{sequence}).  All lasers are switched off during the wait time. Some variability in the measured heating rate was observed, but in all cases the c.m. mode heating rate was measured to be less than 100 quanta/s.

For the data shown in the figures of the main text, the detunings $\Delta$ and Rabi frequencies $\Omega_1$, $\Omega_2$ for EIT cooling are listed in Table \ref{tab:settings}. 
\begin{table}[ht]
    \centering
        \caption{Detunings $\Delta$ and Rabi frequencies $\Omega_1$, $\Omega_2$ that were used for EIT cooling in the measurements shown in the main text.}
    \label{tab:settings}
    \begin{tabularx}{0.5\columnwidth}{l c c c}
    \hline
    \hline
        Figure   & Detuning        & $\Omega_1$    &$\Omega_2$ \\ 
                  & $\Delta$ in MHz & in MHz        &in MHz\\
    \hline
        Fig. 2a)  & 400             & $35.3\pm1.4$  &$35.9\pm1.4$ \\
        Fig. 2b)  & 400             & $35.3\pm1.4$  &$35.9\pm1.4$ \\
        Fig. 3    & 490             & $35.3\pm1.4$  & $35.9\pm1.4$\\
        Fig. 4a)  & 400             &  $34.8\pm1.4$ & $37.1\pm1.5$\\
        Fig. 4c)  & 400             &  $34.6\pm1.4$ & $36.8\pm1.5$\\
    \hline
    \hline
    \end{tabularx}
\end{table}

\section{Normal Modes}
An $N$-ion crystal will support 3$N$ normal modes of oscillation.  For  single-plane crystals in Penning traps, the modes separate into three bands \cite{Wang2013}. As discussed in the main text, there are $N$ drumhead modes that describe ion motion parallel to the magnetic field (z axis). The highest frequency and longest wavelength drumhead mode is the center-of-mass mode at the trap axial frequency $\omega_z$.  The next lower frequency drumhead modes are tilt modes with an effective wavelength of approximately twice the ion crystal diameter.  Because the single-plane crystals lack translational symmetry, the wavelength of a mode is not precisely defined.  However, lower frequency drumhead modes are generally characterized by smaller regions where the ions oscillate in phase, corresponding to effective shorter wavelengths.  The lowest frequency drumhead mode consists of nearest neighbor ions in the center of the crystal oscillating with $\pi$ phase shifts. 

Planar Coulomb crystals have many configurations that are local minima of the potential energy landscape \cite{Bolton1993}.  Crystal rearrangements, which would get averaged over many repetitions of the experiment, might lead to mode frequency instabilities.  We expect larger frequency fluctuations for shorter wavelength modes, because such modes are more sensitive to the local crystal structure than longer wavelength modes whose frequency is almost independent of the crystal configuration. Mode damping due to coupling between the drumhead modes and the low frequency in-plane modes of the crystal (see next paragraph) could also produce apparent frequency fluctuations.

The in-plane modes separate into two dense bands. There are $N$ low frequency $E\times B$ modes and $N$ high frequency cyclotron modes.  Because of Doppler shifts due to the crystal rotation, Doppler cooling of the in-plane modes is not as efficient as with the drumhead modes. With the ion crystal rotation frequency stabilized with a rotating wall potential, modeling of Doppler laser cooling for motion perpendicular to the magnetic field indicates the likelihood of in-plane temperatures $T_{\perp}$ of a few milli Kelvin \cite{Torrisi2016}.  We did not measure $T_{\perp}$ before or after EIT cooling.  In general measuring $T_{\perp}$ is challenging because the thermal motion is very small compared to the coherent rotational motion. Early work \cite{Brewer1988, Gilbert1988},  with three dimensional crystals that did not employ a rotating wall established a bound $T_{\perp}< 10$~mK.  Side-band spectroscopy on a far-detuned stimulated Raman transition could possibly provide a means for careful measurements of in-plane temperatures \cite{Stutter2018}. Currently we do not have the capability of implementing both EIT cooling and a far-detuned stimulated Raman transition.  Future work implementing an in-plane temperature measurement will be important for establishing the impact of EIT cooling on the in-plane modes.

\section{Analytical expression for the bright-state fraction}

As discussed in the main text, we employed spin dephasing induced by coupling ion motion to the spin degree of freedom to measure drumhead mode temperatures.  This technique is well known \cite{Sawyer2012,  Sawyer2014} but not as routinely used for thermometry with trapped ions as sideband thermometry \cite{Lechner2016}.  Sideband thermometry uses motionally generated sidebands on a narrow optical or far-detuned Raman transition for measuring temperatures. It has the nice feature that the measurement has very little background at low temperatures, because there is no motion-subtracting sideband in the motional ground state.  Sideband thermometry with $^9$Be$^+$ requires a far-detuned stimulated Raman transition between the ground state levels, which are separated by 124 GHz.  As discussed in the previous section, we do not have the capability of implementing both EIT cooling and a far-detuned stimulated Raman transition. Instead we employed the spin dephasing for measuring ion temperatures. 

In contrast to sideband thermometry, the spin-dephasing technique is sensitive to zero-point fluctuations and has a background due to induced spin-spin interaction.  Formulas for extracting temperatures with this technique were derived previously in the limit that the induced spin-spin interaction could be neglected (valid for $\bar{n} \gg 1$) \cite{Sawyer2012}.  Below we derive formulas that include the induced spin-spin interaction.  

In the Lamb-Dicke regime, the interaction of the ions with the ODF lasers is described by the Hamiltonian 
\begin{equation}
 \hat{H}=F \cos(\mu_r t)\sum_j \hat{z}_j \hat{\sigma}_j^z,
 \label{eqn:odf_ham_supp}
\end{equation}
\noindent as mentioned in the main text. Here, the position operator for each ion is time-dependent, and can be expanded in terms of the drumhead modes as 
\begin{equation}
    \hat{z}_j(t) = \sum_n \sqrt{\frac{\hbar}{2M\omega_n}}\mathcal{M}_{jn}
    \left( \hat{a}_n e^{i\omega_n t} + \hat{a}_n^\dag e^{-i\omega_n t} \right),
\end{equation}

\noindent where $\omega_n$ is the angular frequency, and $\hat{a}_n,\hat{a}_n^\dag$ the annihilation and creation operators for the normal mode $n$. The matrix element $\mathcal{M}_{jn}$ is the amplitude of the mode $n$ at ion $j$. The ions each have identical mass $M$.

The effect of the spin-echo pulse can be viewed as a change in the sign of $F$ in the second arm of the ODF sequence. Further, an arbitrary phase offset $\phi_0$ can be introduced in the second-arm of the ODF sequence, which modifies the lineshape. These features can be accounted for by the replacement $\cos(\mu_r t)\rightarrow g(t)$, where 
\begin{equation}
    g(t) = \begin{cases}
\cos(\mu_r t),   &  t<\tau      \\       
0,  &   \tau < t < \tau + t_\pi \\
-\cos(\mu_r t + \phi_0), &   \tau + t_\pi < t < 2 \tau + t_\pi.
\end{cases}
\end{equation}

The propagator $\hat{U}(t)$ associated with $\hat{H}$ can be computed exactly, and can be factorized as $\hat{U}(t)=\hat{U}_\text{SM}(t) \times \hat{U}_\text{SS}(t)$, where the spin-motion propagator $\hat{U}_\text{SM}(t)$ and the spin-spin propagator $\hat{U}_\text{SS}(t)$ are given by 
\begin{eqnarray}
    &&\hat{U}_\text{SM}(t) = \prod_n \exp \left[ \sum_j \left( \alpha_{nj}(t)\hat{a}_n^\dag - \alpha_{nj}^* (t) \hat{a}_n \right) \hat{\sigma}_j^z \right],\nonumber\\
    &&\hat{U}_\text{SS}(t) = \exp \left[-i\sum_{i\neq j} J_{ij}(t)\hat{\sigma}_i^z \hat{\sigma}_j^z \right].
    \label{eqn:propagator}
\end{eqnarray}

Here, the coupling constants are given by 
\begin{widetext}
\begin{eqnarray}
    &&\alpha_{nj}(t) = -iF\mathcal{M}_{jn}\sqrt{\frac{\hbar}{2M\omega_n}} \int_0^t dt' g(t') e^{-i\omega_n t'},\nonumber\\
    &&J_{ij}(t) = \text{Im} \sum_n F^2\mathcal{M}_{in}\mathcal{M}_{jn} \frac{\hbar}{2M\omega_n} \int_0^{t} dt_1 \int_0^{t_1} dt_2 g(t_1) g(t_2) e^{i\omega_n(t_2-t_1)}.
\end{eqnarray}
\end{widetext}

We evaluate these expressions within the rotating-wave approximation that $\lvert \delta_n \rvert = \lvert \mu_r-\omega_n \rvert \ll \mu_r +\omega_n$, where the frequency $\delta_n = \mu_r-\omega_n$ is the detuning of the ODF difference frequency from mode $n$. Specifically, at time $2\tau+t_\pi$, the expressions evaluate to 
\begin{widetext}
\begin{eqnarray}
    &&\alpha_{nj}(2\tau+t_\pi) = -\frac{F\mathcal{M}_{jn}}{2\delta_n}\sqrt{\frac{\hbar}{2M\omega_n}}
    \left( e^{-i\delta_n \tau} + e^{-i\phi_0}e^{-i\delta_n (\tau+t_\pi)} - e^{-i\phi_0}e^{-i\delta_n(2\tau+t_\pi)} - 1 \right)\nonumber\\
    &&J_{ij}(2\tau+t_\pi) = \sum_n \frac{F^2\mathcal{M}_{in}\mathcal{M}_{jn}}{4\delta_n^2}
    \frac{\hbar}{2M\omega_n}
    \left( 2\delta_n \tau + \sin(\delta_n(2\tau+t_\pi)+\phi_0) + \sin(\delta_n t_\pi + \phi_0) \right. \nonumber\\
    &&\hphantom{J_{ij}(2\tau+t_\pi) = \sum_n \frac{F^2\mathcal{M}_{in}\mathcal{M}_{jn}}{4\delta_n^2}
    \frac{\hbar}{2M\omega_n}} 
    \left. - 2\sin(\delta_n \tau) - 2\sin(\delta_n(\tau+t_\pi) + \phi_0) \right).\nonumber\\
\end{eqnarray}
\end{widetext}
For $\mu_r$ close to the c.m.\ mode ($n=1$) frequency, here denoted $\omega_1$, the contribution of the other modes is negligible because of the large detunings $\delta_n$ for $n\neq 1$. As a result, $\alpha_{nj}\approx 0$ for $n\neq 1$ and the symmetric coupling of the c.m.\ mode to all the ions results in $\alpha_{1j}(2\tau+t_\pi) \equiv \alpha$ and $J_{ij}(2\tau+t_\pi)\equiv J$, independent of the ion numbers $i,j$, with expressions 
\begin{widetext}
\begin{eqnarray}
    &&\alpha = -\frac{F}{2\sqrt{N}\delta_1}\sqrt{\frac{\hbar}{2M\omega_z}}
    \left( e^{-i\delta_1 \tau} + e^{-i\phi_0}e^{-i\delta_1 (\tau+t_\pi)} - e^{-i\phi_0}e^{-i\delta_1(2\tau+t_\pi)} - 1 \right)\nonumber\\
    &&J = \frac{F^2}{4N\delta_1^2} \frac{\hbar}{2M\omega_z}
    \left( 2\delta_1 \tau + \sin(\delta_1(2\tau+t_\pi)+\phi_0) + \sin(\delta_1 t_\pi + \phi_0) - 2\sin(\delta_1 \tau) - 2\sin(\delta_1(\tau+t_\pi) + \phi_0) \right).\nonumber\\
\end{eqnarray}
\end{widetext}

The lineshapes in Fig.\ 2(a) and 2(b) of the main text are obtained respectively with phase offsets of $\phi_0 = 0$ and $\pi$. The bright-state fraction $P(\ket{\uparrow}) = \sum_j 1/2 (1-\langle \hat{\sigma}_j^z \rangle)$, where $\langle \hat{\sigma}_j^z \rangle$, the population difference \emph{after} the final $\pi/2$ pulse, is equal to $-\langle \hat{\sigma}_j^x \rangle$ \emph{before} that $\pi/2$ pulse.     

With the modes initially in thermal states characterized by mean occupations $\bar{n}_n$, and the spins initialized along the $x$ direction, the evolution of any observable can be computed using the propagator $\hat{U}(t)$ whose form is detailed in Eq.\ (\ref{eqn:propagator}) \cite{WallSafaviSB}.  Specifically, the expression for $\langle \hat{\sigma}_j^x \rangle$ evaluates to 
\begin{equation}
    \langle \hat{\sigma}_j^x \rangle = 
    \left\{ \prod_{i\neq j}\cos(4J_{jk})\right\}
    \exp \left[ -2 \sum_{n} \lvert \alpha_{nj} \rvert^2 \left( 2\bar{n}_n + 1 \right) \right],
\end{equation}

\noindent which for $\mu_r$ close to the c.m. mode becomes independent of $j$ and reduces to 
\begin{equation}
    \langle \hat{\sigma}^x \rangle = 
    \left(\cos(4J)\right)^{N-1}
    \exp \left[ -2 \lvert \alpha \rvert^2 \left( 2\bar{n}_1 + 1 \right) \right].
\end{equation}

Finally, accounting for free-space scattering at a rate $\Gamma$ for the time the ODF lasers are turned on, we arrive at the expression, Eq.\  (1) of the main text, for the bright state fraction at the end of the ODF sequence. 

\bibliographystyle{apsrev4-1}
\bibliography{EITcooling}